\documentclass[twocolumn,prb,showpacs,preprintnumbers,amsmath,amssymb,floatfix]{revtex4}

\usepackage{graphicx}
\usepackage{dcolumn}
\usepackage{bm}

\newcommand{\unite}[1]{\text{$\mathrm{ #1 }$}}

\newcommand{\micrometer}{\text{{$\mu$}m}}
\newcommand{\E}[1]{\times 10^{#1}}

\newcommand{\Jm}[1]{\unite{J/m^{#1}}}
\newcommand{\Py}{${\rm{Fe}_{20}\rm{Ni}_{80}}$}

\newcommand{\DW}{DW}

\newcommand{\wstar}{w^*}
\newcommand{\wzero}{w_0}
\newcommand{\Cw}{d_0}
\newcommand{\Rw}{d_1}
\newcommand{\Cl}{S_0}

\newcommand{\spinSEM}{spin-SEM}
\newcommand{\SEMPA}{SEMPA}

\newcommand{\Aexch}{A_{\rm{exch}}}
\newcommand{\Ms}{M_{\rm{s}}}

\begin{document}

\title{Magnetic domain walls in constrained geometries}

\author{P.-O. Jubert}
 \email{pju@zurich.ibm.com}
\author{R. Allenspach}
\author{A. Bischof}

\affiliation{IBM Research, Zurich Research Laboratory, CH-8803
R\"uschlikon, Switzerland}

\begin{abstract}

Magnetic domain walls have been studied in micrometer-sized \Py\
elements containing geometrical constrictions by spin-polarized
scanning electron microscopy and numerical simulations. By
controlling the constriction dimensions, the wall width can be
tailored and the wall type modified. In particular, the width of a
$180^{\circ}$ N\'{e}el wall can be strongly reduced or increased
by the constriction geometry compared with the wall in
unconstrained systems.

\end{abstract}

\maketitle

\vskip 0.5in

For almost a century, the width of magnetic domain walls (\DW\/s)
has been believed to be determined by material properties only.
However, recent investigations on \DW\/s in nanometer-scale
systems have revealed new physical properties due to the
geometrical confinement of the magnetization. A reduction of the
Bloch wall width has been predicted \cite{Bruno1999} and observed
\cite{Pietzsch2000PRL} in nanometer-sized constrictions. This
effect is thought to be the origin of the large magneto-resistance
measured in nanocontacts, and explained by ballistic transport
through a narrow \DW\ pinned within the contact.
\cite{Garcia1999,Chopra2002,Ruester2003} Furthermore, domain walls
are now being investigated as tiny individual magnetic objects
that can be manipulated in view of their potential for application
in novel magnetic logic or memory devices.
\cite{Allwood2002Science} Of interest in this field are the
possibilities of pinning \DW\/s at constrictions and of displacing
them using a magnetic field \cite{Ono1999Science} or an applied
current.
\cite{Gan2000IEEE,Grollier2002JAP,Grollier2003APL,Vernier2004,Yamaguchi2004}
For all these phenomena, the key parameter is the magnetic
structure of the domain walls. For a basic understanding as well
as for potential applications, it is important to gain
quantitative insight into how \DW\ properties can be modified via
the geometry.

The prediction of \DW\ narrowing in a constriction was based on a
ferromagnetic model system containing a planar Bloch wall.
\cite{Bruno1999} Because dipolar contributions in the constriction
were neglected, the problem was one-dimensional and could be
solved analytically. The vast majority of small elements, however,
exhibits \DW\/s of N\'{e}el type, with a nonvanishing
magnetization component perpendicular to the wall. In these walls,
the dipolar energy determines the wall profile to a large extent,
and hence the problem is more intricate.

In this paper, we investigate N\'{e}el-type walls in elements
containing constrictions of controlled dimensions. The
experimental results obtained by scanning electron microscopy with
spin analysis (\spinSEM\ \cite{Koike1984JAP} or \SEMPA\
\cite{Scheinfein1990}) are compared with micromagnetic
simulations. We demonstrate how the wall properties can be tuned
both by the element size and the constriction dimensions.
Constraining a \DW\ in a micrometer-sized element strongly reduces
the wall width compared with the width in an infinite film. By
appropriately tuning the constriction dimensions, the N\'{e}el
wall width can further be decreased, or alternatively, increased
until the wall splits into two separate walls.

The constrictions were fabricated in thin, micrometer-sized
rectangular elements by using electron-beam lithography and Ar dry
etching of \Py\ thin films. These films were produced by sputter
deposition on $\rm{SiO}_2$/Si(100), resulting in (111)-textured
\Py\ films with a grain size of about 2 nm diameter, as determined
from x-ray measurements. The crystalline anisotropy of the film
was measured to be a superposition of a uniaxial term ${K_{\rm{u}}
= 160 \unite{\; \Jm{3}}}$ and a cubic term ${K_{\rm{1}} = -150
\unite{\; \Jm{3}}}$. The anisotropy is so low that it has no
influence on the magnetic patterns in this study, as verified by
micromagnetic simulations. All simulations were carried out using
the OOMMF code. \cite{oommfREF} The material parameters used for
the simulations are the commonly employed values for \Py\/, an
exchange constant $ \Aexch= 13\E{-12} \; \Jm{} $ and a saturation
magnetization $M_s = 800$ kA/m. The discretization cell size was 5
nm.

\begin{figure}[ht]
  \begin{center}
  \leavevmode
  \includegraphics[width=80mm]{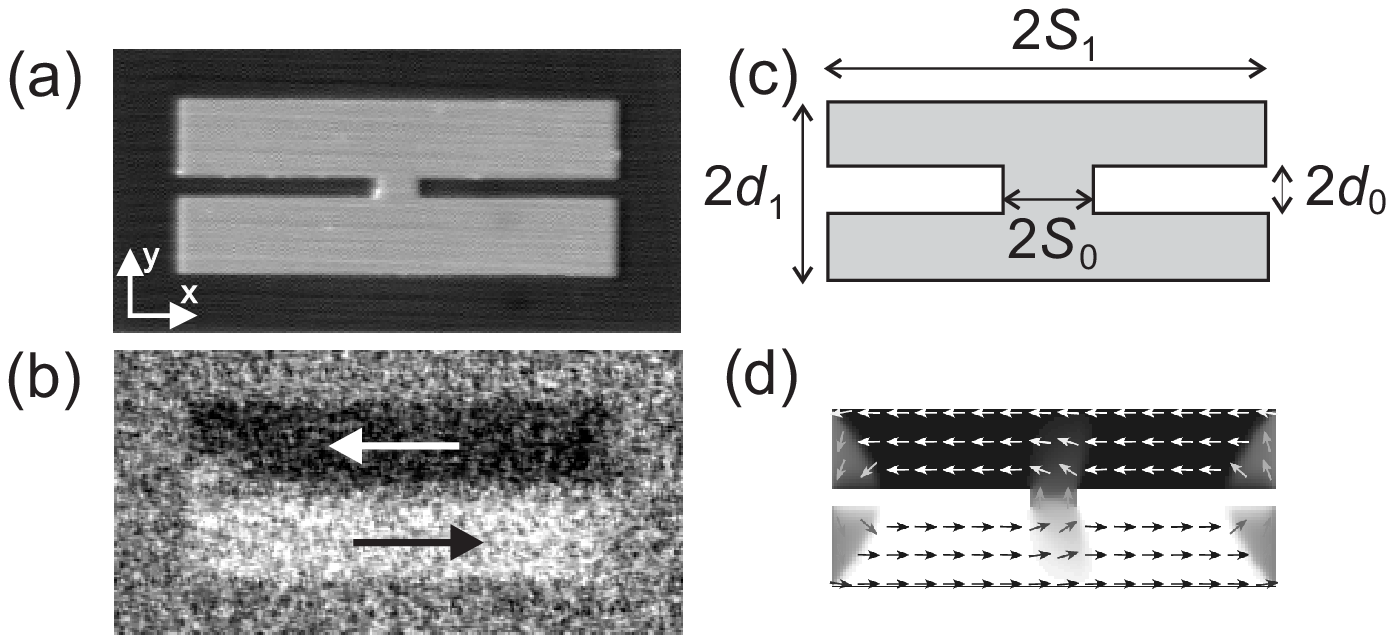}
  \end{center}
  \vspace*{-5mm}
  \caption{(a) Topographic image of a $ 10\unite{\; \micrometer}\times
  4\unite{\; \micrometer}\times 7.5\unite{\; nm}$ \Py\ element
  with constriction dimensions $ \Cw=225 \unite{\; nm}$ and $ \Cl=500\unite{\; nm}$.
  (b) Corresponding magnetic configuration after
  ac demagnetization as determined by spin-SEM.
  The arrows indicate the magnetization direction.
  (c) Schematic of the rectangular magnetic element with constriction.
  (d) Magnetic configuration calculated for an element having the same dimensions.
  }
  \label{Fig-spinSEM}
\end{figure}

The magnetic configurations of the elements were investigated in
our spin-SEM setup. \cite{Allenspach1994JMMM} Topography and
magnetization distribution are  determined simultaneously and with
a lateral resolution of 20 to 30 nm. Thus, the technique is
ideally suited for measuring \DW\ profiles.
\cite{Scheinfein1991,Berger1992,Vaz2003} The patterned sample was
cleaned \textit{in-situ} by gentle Ne sputtering to remove surface
contaminants. Eventually, 1--2 monolayers of Fe were deposited
\textit{in-situ} onto the \Py\ sample. This enhances the electron
spin polarization without affecting the magnetization
configuration of the element. \cite{vanZandt1991}

The shape of the micrometer-sized elements has been chosen such as
to facilitate the pinning of a $180^{\circ}$ N\'{e}el wall inside
the constriction. Because of the predominant magnetostatic
contribution in \Py\/, the lowest energy state is a two-domain
configuration separated by a $180^{\circ}$ N\'{e}el wall
positioned in the constriction, see Fig.\ \ref{Fig-spinSEM}. This
configuration is obtained experimentally after ac demagnetization.
Without constriction, a single domain prevails. We restrict the
film thickness to $\leqslant$ 20 nm to ensure that the N\'{e}el
walls are homogeneous throughout the element thickness.

\begin{figure}[t]
  \begin{center}
  \leavevmode
  \includegraphics[width=80mm]{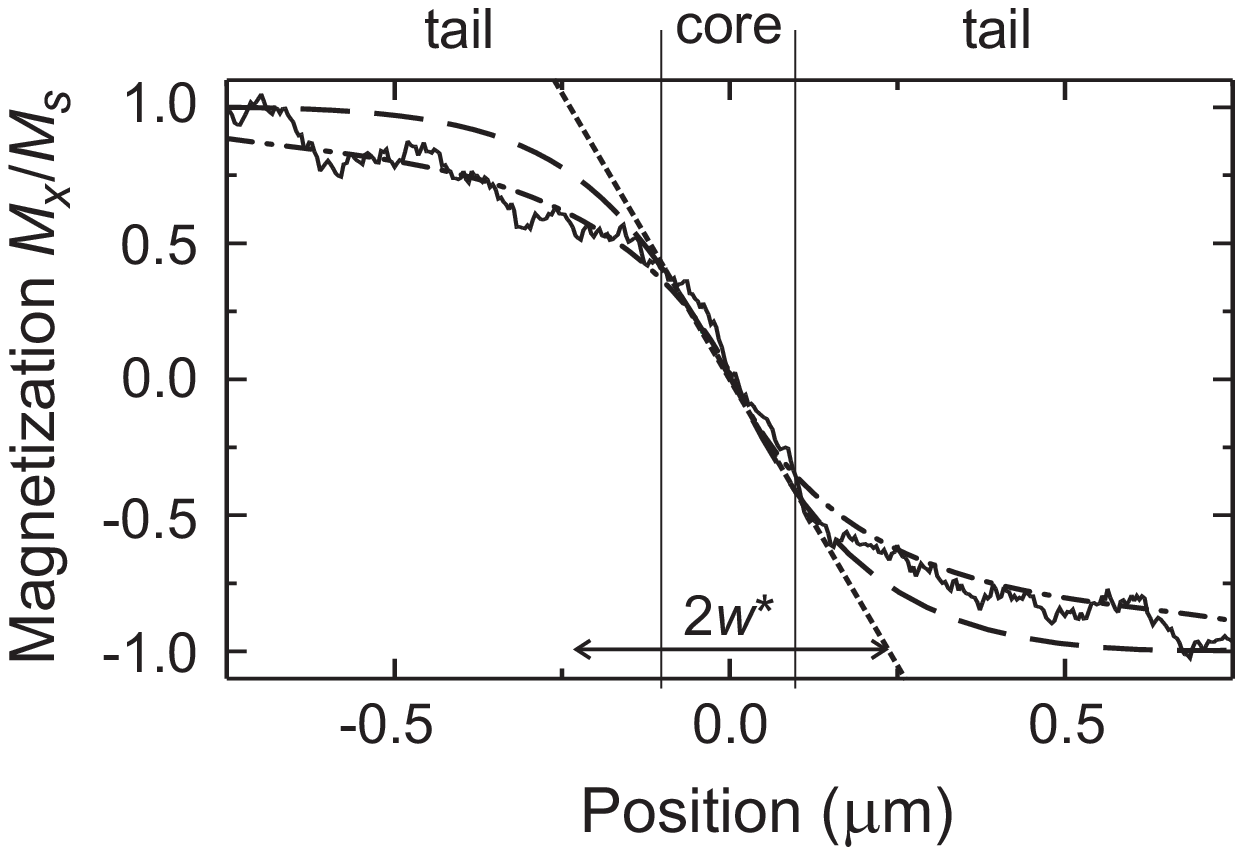}
  \end{center}
  \vspace*{-5mm}
  \caption{Domain-wall profile measured for a
  $ 10\unite{\; \micrometer}\times 4\unite{\; \micrometer}\times 7.5\unite{\;
  nm}$
  \Py\ element with a constriction size of $ \Cw=225\unite{\; nm}$ and
  $\Cl=500\unite{\; nm}$. The experimental profile is
  compared with the calculation (dash-dotted line) and an hyperbolic
  tangent function (dashed line). The profile is averaged over
  ${80\%}$
  of the constriction length $2 \Cl $.
  }
  \label{Fig-DWprofile}
  \end{figure}

The profile across the wall pinned in the constriction of Fig.\
\ref{Fig-spinSEM} is shown in Fig.\ \ref{Fig-DWprofile}. Excellent
agreement with the calculated profile is found. The typical
features of a symmetrical N\'{e}el wall can be identified: The
magnetization mainly rotates in the core of the wall, which can be
approximated by a hyperbolic tangent function. In the outer
regions, the profile shows a slower magnetization rotation and
hence deviates from the $\tanh$ function. These long tails are
typical for the N\'{e}el wall, and have been identified in thin
films earlier. \cite{Berger1992} We quantify the wall by defining
a mean \DW\ width $ \wstar $, which is the inverse slope of the
curve, fitted linearly in the range $ -0.4 < M/\Ms < 0.4 $.

\begin{figure}[t]
  \begin{center}
  \leavevmode
  \includegraphics[width=80mm]{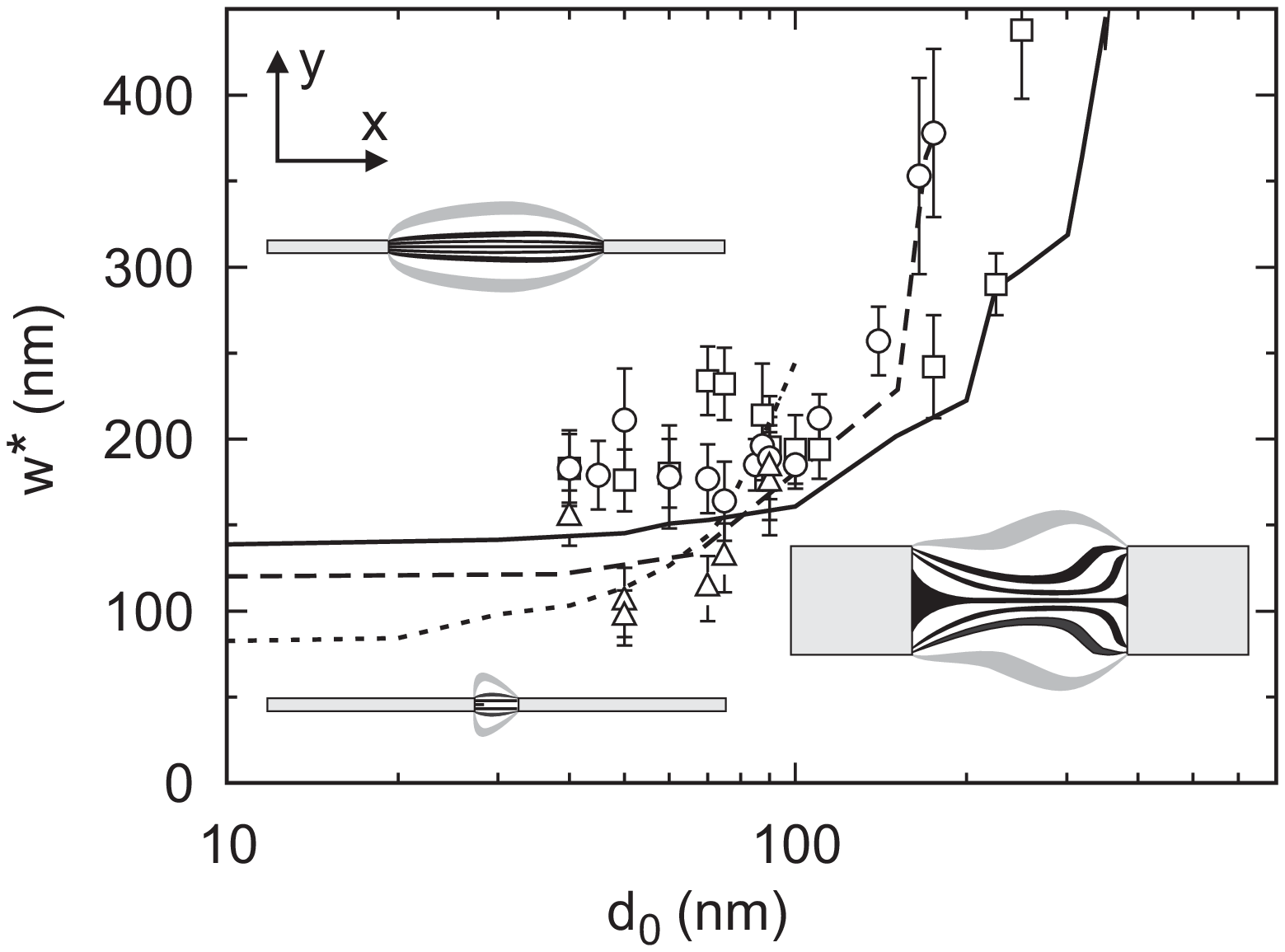}
  \end{center}
  \vspace*{-5mm}
  \caption{Evolution of the average \DW\ width $ \wstar $  vs.\ constriction dimension
  $ \Cw $ in \Py\   with $ \Cl $ as parameter ranging from
  100 nm (triangles, dotted line) to 250 nm (circles, dashed line) and 500 nm
  (squares, solid line);   element size:
  $ 10\unite{\;\micrometer}\times 4\unite{\;\micrometer}\times 7.5\unite{\; nm} $.
  Dots are experimental values and lines the calculated \DW\ width.
  Insets show constant levels of calculated domain configurations.
  $ M_x = 0, 0.25, 0.5 \Ms $ correspond to the core of the domain wall (black)
  and $ M_x = 0.75 \Ms $ to its tails (grey).
  $ \Cw=30\unite{\; nm}$ and $ \Cl=100\unite{\; nm}$
  (lower left), $ \Cw=30\unite{\; nm}$ and
  $ \Cl=1\unite{\; \micrometer}$ (upper left), $ \Cw=225\unite{\;
  nm}$ and $ \Cl=500\unite{\; nm}$ (right).
  }
  \label{Fig-DWw-Const}
  \end{figure}

To evaluate the influence of the geometry on the wall
profile, the dimensions of the constriction have been
varied systematically. Figure \ref{Fig-DWw-Const} presents
the evolution of $ \wstar $as a function of $ \Cw $ and $
\Cl $ for magnetic elements of constant dimensions, $
10\unite{\; \micrometer} \times 4\unite{\; \micrometer}
\times 7.5\unite{\; nm}$. For $\Cl=500\unite{\; nm}$ and $
\Cw<100\unite{\; nm}$, the average \DW\ width is almost
constant. However, a steep increase of $ \wstar $occurs as
$ \Cw $ becomes larger than $ 100\unite{\; nm}$. A similar
variation is observed when $ \Cl=250\unite{\; nm}$, but the
increase of $ \wstar $at large $ \Cw $ values is steeper.
Finally, a clear reduction of the wall width is found for
even smaller constrictions with $ \Cl=100\unite{\; nm}$ and
$ \Cw < 100\unite{\; nm}$. The numerical simulations
reproduce the experimental trend well and allow us to
explain the observed wall-width variations qualitatively.

Consider an unconstrained $180^{\circ}$ N\'{e}el wall in an
extended film. Its total energy is given by the wall energy
density multiplied by the total length. The wall width results
from the minimization of the energy density, and is solely
determined by the material parameters. A \DW\ located in a
constriction, however, is affected by the constriction geometry.
At the constriction edges, the magnetization is forced to lie
parallel to the sides to minimize the surface dipolar energy, and
hence the wall width locally corresponds to $ \Cw $. To minimize
its total energy in such a geometry, the \DW\ deforms in the
constriction area, i.e., the local wall width depends on the
distance from the constriction center, and the wall adopts a
two-dimensional shape. We consider the two limiting cases of small
and large $ \Cw $. The relevant length scale to compare with is
the wall width without constriction, $ \wzero $, which deviates
from the wall width in an extended film as discussed below.

For 2$ \Cw $ $<$ $ \wzero $, an outward deformation of the \DW\
results (see insets on the left in Fig.\ \ref{Fig-DWw-Const}). For
large $ \Cl $, this deformation is restricted to the edge region
and does not affect the \DW\ profile in the center of the
constriction. The measured average width is then almost constant
and $ \wstar \simeq \wzero $. For smaller $ \Cl $, however, the
center wall width cannot be kept at $ \wzero $. It costs too much
exchange energy to let the local wall width vary rapidly along the
constriction. Therefore, the wall at the center reduces its width
to less than $ \wzero $. Accordingly, below a threshold value of $
\Cl \simeq 250\unite{\; nm}$, a reduction of $ \wstar $is observed
with decreasing $ \Cl $, which becomes most pronounced for small $
\Cl $.

For 2$ \Cw $ $>$ $ \wzero $ $\backsimeq 200$ nm, the wall
width at the constriction edges expands and the wall
deforms inward (see inset on the right in Fig.\
\ref{Fig-DWw-Const}). The same energy arguments as above
apply: For small $ \Cl $, the wall at the center stretches,
resulting in an increase of the measured $ \wstar $. The
larger $ \Cl $ the stronger the confinement of the wall
deformation to the edge region, and consequently the
increase of $ \wstar $is postponed to larger $ \Cw $
values.

\begin{figure}[ht]
  \begin{center}
  \leavevmode
  \includegraphics[width=70mm]{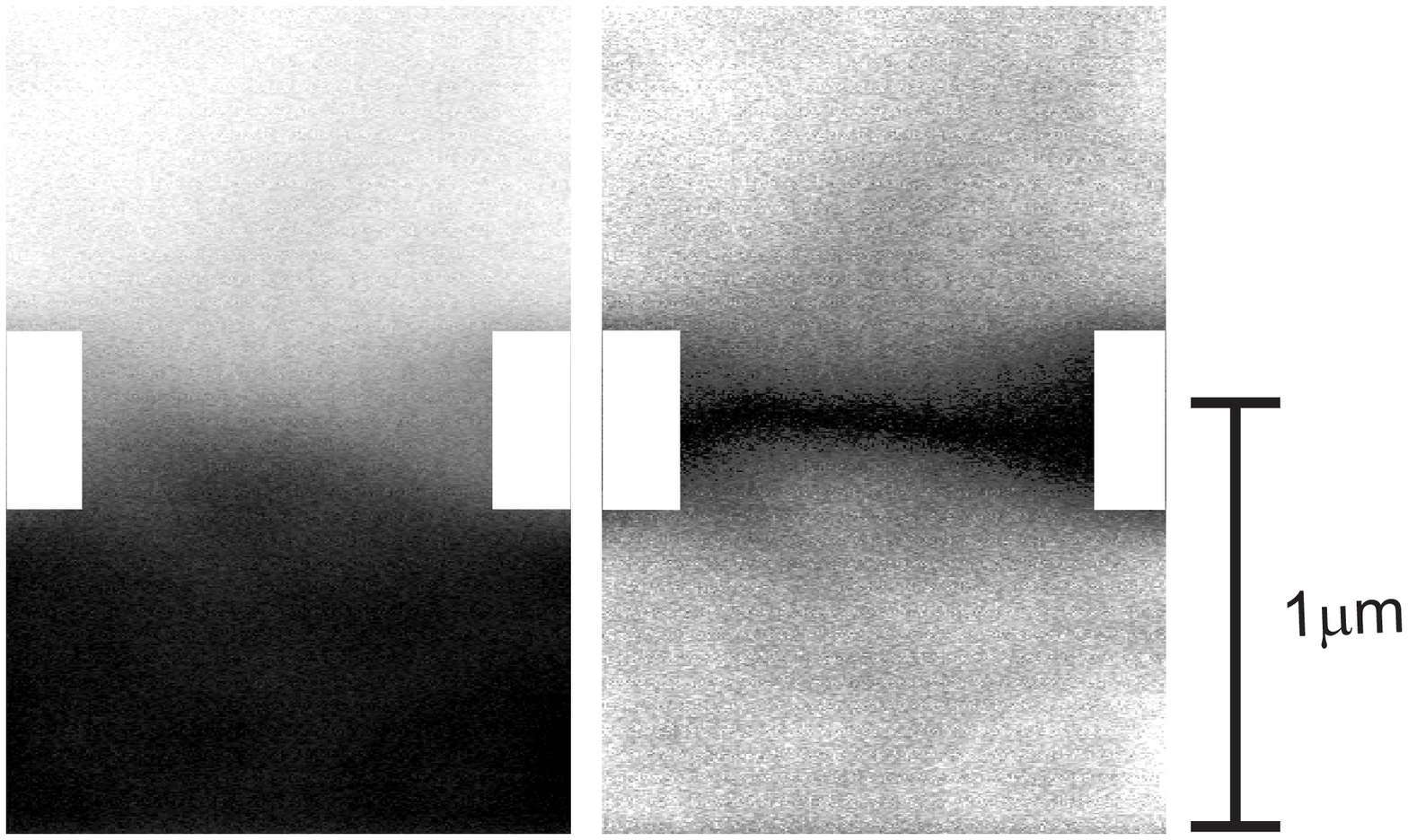}
  \end{center}
  \vspace*{-5mm}
  \caption{DW configuration measured in a constriction with
  $ \Cw=225\unite{\; nm}$ and $ \Cl=500\unite{\; nm}$; element
  size:   $ 10\unite{\; \micrometer}\times 4\unite{\; \micrometer}\times 7.5\unite{\; nm}$.
  Grey-scale images correspond to the magnetization component
  parallel (left) and perpendicular (right) to the wall.
  The wall is asymmetric with respect to the left and right
  constriction edges. The position of the constriction
  has been determined from the topographic image.
  }
  \label{Fig-DWconfig}
  \end{figure}

Figure \ref{Fig-DWconfig} illustrates the complexity of the
\DW\ in the constriction. The wall is two-dimensional and
asymmetric, in good agreement with the calculated
configuration. In the constriction region, the
magnetization has to undergo a $180^{\circ}$ rotation. As
exchange forces the spins to remain as closely aligned with
each other as possible, the $180^{\circ}$ turn happens over
a larger distance in the outer part of the turn. This
explains the asymmetry of the \DW\ configuration. We
emphasize that the complex two-dimensional structure of a
\DW\ is not restricted to the N\'{e}el wall. We have also
simulated Bloch walls and find similar features:\ bent
walls and complex two-dimensional patterns. The assumption
that a Bloch wall can be treated as a one-dimensional
object \cite{Bruno1999} is only justified for $ \Cw \ll \Cl
$. Otherwise, dipolar contributions force the Bloch wall
configuration to adopt a N\'{e}el-type arrangement.

\begin{figure}[ht]
  \begin{center}
  \leavevmode
  \includegraphics[width=70mm]{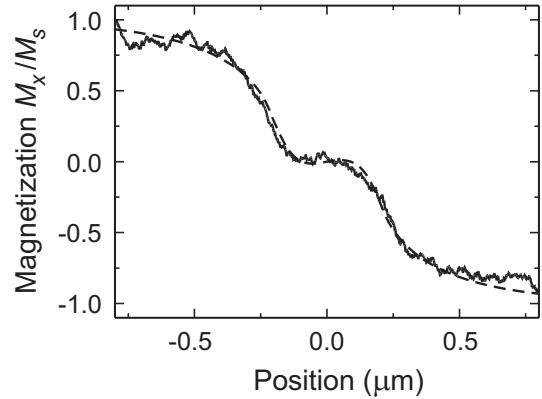}
  \end{center}
  \vspace*{-5mm}
  \caption{Wall profile determined in a wide constriction with
  $ \Cl=250\unite{\; nm} $ and $ \Cw=250\unite{\; nm}$ showing the splitting
  of the wall into two individual $90^{\circ}$ walls.
  Experimental data are shown as a solid line, the simulation as a dashed line; element
  dimensions: $ 10\unite{\; \micrometer}\times 4\unite{\; \micrometer}\times
  7.5\unite{\; nm} $.
  }
  \label{Fig-doubleDW}
  \end{figure}

Figure \ref{Fig-DWconfig} also reveals the presence of a
small intermediate domain nucleated at one constriction
edge. This results from a strong preference for low-angle
\DW\/s in the case of N\'{e}el walls: Energy can be gained
by replacing the $180^{\circ}$ domain wall by two
$90^{\circ}$ walls and an intermediate domain. The
splitting into two $90^{\circ}$ walls is local at first and
will not affect the average \DW\ profile significantly. But
as $ \Cw $ approaches $ \Cl $, this domain expands to the
constriction center, leading to the complete separation of
the $180^{\circ}$ wall into two $90^{\circ}$ walls. This is
experimentally visible in the average wall profile, as
shown in Fig.\ \ref{Fig-doubleDW} for a constriction of
dimensions $ \Cl=\Cw=250\unite{\; nm} $. Thus, an
additional condition needs to be fulfilled to avoid wall
splitting and to pin a narrower $180^{\circ}$ \DW\ in a
constriction. Apart from $ \Cl $ being small, also $\Cw <
\Cl$ is required.

\begin{figure}[t]
  \begin{center}
  \leavevmode
  \includegraphics[width=60mm]{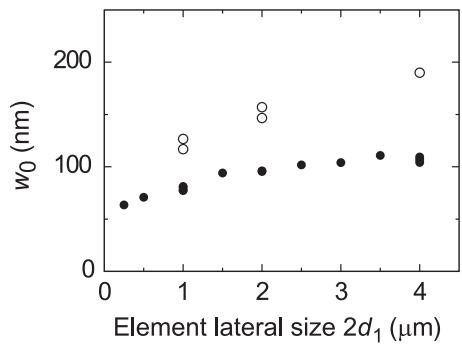}
  \end{center}
  \vspace*{-5mm}
  \caption{Evolution of the average \DW\ width $ \wstar $  as a function of the
  rectangle's lateral size $ 2 \Rw $, with
  experimental results (empty circles) and calculations (filled circles).
  While a constriction was needed to pin a $180^{\circ}$ N\'{e}el wall its
  dimensions are such that $ \wstar $ equals $ \wzero $ (see Fig. \ref{Fig-DWw-Const}):
  $ \Cl = $ 250 nm and $ \Cw =$ 50 nm. The element thickness is 20 nm.
  }
  \label{Fig-DWw-vs-Lt}
  \end{figure}

Having considered the effect of the constriction dimensions on the
\DW\ width, let us discuss the influence of the element size. In
an infinitely extended 10-nm-thick \Py\ film, the N\'{e}el wall
width has been determined to be on the order of 100 nm,
\cite{Suzuki1968,Wong1996} with tails extending several
micrometers beyond the core. In our small magnetic element,
however, the tails are limited by the element width $ 2 \Rw $,
leading to modified profile and wall width. Compared with the
unconstrained extended film, the magnetostatic charges need to be
redistributed within the N\'{e}el wall. This rearrangement of
charges is rather complex and has not yet been described
theoretically. \cite{Holz1969}  Figure \ref{Fig-DWw-vs-Lt} shows
the variation of the wall width $ \wzero $ determined for
rectangular elements of varying lateral size. Both, our
experimental results and micromagnetic simulations demonstrate
that a strong reduction of $ \wzero $ occurs when decreasing $ 2
\Rw $, owing to the confinement of the wall in the
micrometer-sized element. Such a dependence on the
magnetic-element dimensions is specific of the N\'{e}el wall and
is not expected to occur for a Bloch wall.

Finally, let us comment on the small discrepancy observed between
the experimental and calculated \DW\ widths at small $ \Cw $. This
is reminiscent of other experimental \DW\ profile measurements in
which the \DW\ width was found to be much larger than the
simulated values. \cite{Vaz2003} It was proposed that the
magnetocrystalline anisotropy is smaller and/or the exchange
constant larger than assumed in the calculations. Anisotropy can
be ruled out in our case. The fact that the discrepancy arises
only at small $ \Cw $, at which exchange is predominant,
corroborates that exchange might be underestimated in the
simulations.

In conclusion, we have investigated the configuration of
$180^{\circ}$ N\'{e}el walls pinned in patterned
constrictions, both experimentally and by micromagnetic
simulations. Isolated domain walls present a complex
configuration which is two-dimensional and asymmetric. We
have further demonstrated that their average width can be
tuned by the dimensions of both the element and the
constriction. Because of the extended tails in the N\'{e}el
walls, the wall width in micrometer-sized elements can be
strongly reduced,  compared with the infinitely extended
thin film. The width decreases when the lateral size of the
magnetic element $ 2 \Rw $ decreases. In addition, by
tuning the constriction dimensions, the N\'{e}el-wall width
can also be modified continuously. Owing to significant
dipolar contributions at the constriction edges, the domain
wall can be stretched to wide profiles for large
constriction widths $ 2\Cw $. For small $ \Cw $, on the
other hand, the domain wall is strongly confined when the
constriction length $ 2\Cl $ decreases. The ability to
control domain-wall profile properties through geometrical
constraints calls for further investigation of other
domain-wall properties such as the magnetoresistance, or
the displacement of constrained magnetic domain walls.

We acknowledge James Jarratt and Ian McFadyen for providing us
with the \Py\ films, Michel Despont, Hugo Rothuizen, Ute
Drechsler, and Richard Stutz for their contribution in the
patterning processes of the films, and Daniele Mauri and Jerome
Wolfman for their input in the initial phase of the project.


\end{document}